\def\rfr#1{eq. (\ref{#1})}

\def\bm#1{{\mbox{\boldmath$#1$\unboldmath}}}

\def\bar{\begin{eqnarray}}
\def\ear{\end{eqnarray}}

\def\eqi{\begin{equation}}
\def\eqf{\end{equation}}
\def\eqia{\begin{eqnarray}}
\def\eqfa{\end{eqnarray}}
\def\rp#1#2{{#1\over#2}}

\def\lb#1{\label{#1}}

\def\oc2{$\mathcal{O}(c^{-2})$}

\documentclass[page-classic]{epl2}

\title{LARES/WEBER-SAT and the equivalence principle}
\shorttitle{LARES/WEBER-SAT and the equivalence principle}

 \author{L. Iorio\inst{1}}
 \shortauthor{L. Iorio}

 \institute{ \inst{1} Viale Unit$\grave{a}$ di Italia 68, 70125, Bari (BA), Italy
\\tel. 0039 328 6128815
\\e-mail: lorenzo.iorio@libero.it}


\abstract{
It has often been claimed that   the proposed  Earth artificial satellite LARES/WEBER-SAT$-$whose primary goal is, in fact, the measurement of the general relativistic Lense-Thirring effect at a some percent level$-${would allow greatly improving}, among (many) other things, the present$-$day ($10^{-13}$) level of accuracy in testing the equivalence principle as well. Recent claims point towards even two orders of magnitude better, i.e. $10^{-15}$.
In this note we show that such a goal is, in fact, unattainable by many orders of magnitude being, instead, the achievable level $\approx 10^{-9}$.}

\begin{document}

\maketitle

\section{The LAGEOS III/LARES/WEBER-SAT mission}
In 1976 Van Patten and Everitt \cite{Van76a,Van76b} suggested  {measuring} the general relativistic Lense-Thirring node precession \cite{Len18} with  a pair of counter-orbiting spacecraft to be placed in terrestrial polar orbits and endowed with active, drag-free apparatus to counter-act the non-gravitational perturbations.   In 1977-1978 Cugusi and Proverbio \cite{Cug77,Cug78}  proposed {using} the existing  passive geodetic satellite LAGEOS and, more generally, the Satellite Laser
Ranging (SLR) technique to measure the Lense-Thirring effect in
the terrestrial gravitational field with the existing artificial
satellites. In 1986 Ciufolini \cite{Ciu86} put forth a strategy somewhat equivalent to the ones by Van Patten and Everitt and Cugusi and Proverbio involving the launch of a passive, LAGEOS-like satellite$-$named LAGEOS III after the launch of LAGEOS II$-$in an orbit identical to that of LAGEOS (semimajor axis $a=12,270$ km, inclination of the orbital plane to the Earth's equator $i=110$ deg, eccentricity $e=0.0045$), apart from the inclination which should have been equal to $i=70$ deg. The observable originally proposed was the sum of the nodes of LAGEOS and LAGEOS III: the particular orbital configuration of the new spacecraft was chosen to cancel out the aliasing impact of the even zonal harmonic coefficients $J_{\ell}, \ \ell=2,4,...$ of the multipolar expansion of
the Newtonian part of the Earth's gravitational potential which induce node precessions qualitatively identical to the relativistic ones but much larger. Although extensively studied by various groups \cite{Ri89,Iorio02}, such an idea has not yet been implemented. In 1998, {after its} originally proposed eccentricity was augmented by almost one order of magnitude in order to use the perigee as well \cite{Ciu98}, renamed LAser RElativity Satellite (LARES), it was rejected by the Agenzia Spaziale Italiana (ASI, Italian Space Agency). With the name of WEBER-SAT\footnote{In memory of Dr. J. Weber, US Naval Academy (USNA) Class of 1940}, it is currently under examination by the Istituto Nazionale di Fisica Nucleare\footnote{See on the Internet
{http://www.lnf.infn.it/acceleratori/lares/}
and {http://cadigweb.ew.usna.edu/~webersat/}
} (INFN, National Institute of Nuclear Physics).

In addition to the original, by far principal, purpose of measuring the Lense-Thirring effect at some percent level, many other goals of fundamental physics have been recently added to the LARES mission in order to enhance its chances of being finally approved.  In doing so it has gone  too far by often making unrealistic claims.
%
%
 A typical example of such a policy is represented by the alleged possibility \cite{Ciucaz04} of measuring the perigee precession \cite{Lue} induced by the multi-dimensional braneworld model by Dvali, Gabadadze and Porrati (DGP) \cite{Dvali}.  Amounting to $4\times 10^{-3}$ milliarcseconds per year (mas yr$^{-1}$), it has been proven to be undetectable in \cite{Iorio05a}. In such a work motivations mainly concerning the gravitational systematic errors were considered: here we will yield further, simple  arguments related to the non-gravitational perturbations which the perigees of LAGEOS-like satellites are particularly sensitive to. In \cite{Cantone} it has been reported that the INFN team should be able to reduce the impact of the non-gravitational perturbations of thermal origin down to $\approx 10^{-3}$ of the Lense-Thirring effect on LARES. Although not explicitly stated in \cite{Cantone}, let us assume that this will be true for the perigee as well, which is certainly not an easy task to be implemented: the Lense-Thirring perigee precession for the originally proposed LARES orbital configuration amounts to about 30 mas yr$^{-1}$, so that a mismodelled precession of $\approx 10^{-2}$ mas yr$^{-1}$ of non-gravitational origin would be left, i.e. a bias just one order of magnitude larger than the DGP effect itself.
\section{The equivalence principle}
 Another example is the equivalence principle. Indeed, in several works \cite{Ciu04,Ciu05,Della05,Smi1,Smi2} it has been recently claimed that LARES/WEBER-SAT would be able to greatly improve the accuracy level in testing such a cornerstone of general relativity and of all other competing metric theories of gravity. E.g., in \cite{Della05} we find: ``An additional physics goal of LARES is the
improvement of the limits on the violation of the Einstein Equivalence Principle{.}'' More precisely, in \cite{Ciu04} it is explicitly written: ``LARES would improve, by about two orders of magnitude, the accuracy in testing the equivalence principle [...]''. Now, since the present$-$day level of accuracy in testing it is $10^{-13}$ \cite{Wil06},
 it must be argued that Ciufolini  is claiming in \cite{Ciu04} that a level of $10^{-15}$ is to be expected from the successful implementation of the LARES mission.  Unfortunately, the authors of such claims nowhere explicitly explain how they would achieve such notable goals.

 {By contrast}, the situation is quite different, and by many orders of magnitude. The possibility of testing the equivalence principle with artificial Earth satellites of different compositions was tackled in \cite{Bla01,Iorio04,Nob07}. In \cite{Iorio04} {just} the originally proposed configuration of LARES ($a=12,270$ km, $i=70$ deg) was  examined: the existing LAGEOS satellite would be used in conjunction with LARES/WEBER-SAT which would substantially differ from it by composition, weight and manufacturing. Both the orbital periods $T$ and the secular precessions of the nodes  were considered{,} finding that the choice of the orbital periods would be better by one order of magnitude. In the former case, one should measure, after many revolutions, the difference of time spans which are multiple $N$ of the orbital periods $\Delta T_N=T_N^{(2)} - T_N^{(1)}$  of the {pair} of satellites, here denoted with the superscripts $(1)$ and $(2)$, orbiting the Earth along paths with almost the same semimajor axes $a$, {to be measured from SLR data as well}
  \eqi a^{(1)}\equiv a,\ a^{(2)}=a+d.\eqf   Indeed, the equivalence principle-violating parameter $\Delta\psi$ can be written in terms of $\Delta T_N$ as  \cite{Iorio04}
\eqi\Delta\psi_N = \rp{\rp{\sqrt{GM}}{N\pi}\Delta T_N -3d\sqrt{a} }{3d\sqrt{a}+2\sqrt{a^3}}.\lb{ciufep}\eqf
The major limiting factor is mainly represented by the fact that, due to the unavoidable orbital
injection errors, it would not be possible to insert the new satellite in an orbit with exactly the same semimajor axis of LAGEOS and that
such difference $d$ could be known only with a finite precision. For  $d\approx 5$ km and an uncertainty\footnote{Recall that both the existing LAGEOS and the proposed LARES are not endowed with any satellite-to-satellite tracking system or any other apparatus apart from retroreflectors.} $\delta d\approx 1$ cm the obtainable accuracy would be $10^{-9}$ only, i.e. four orders of magnitude less than the present$-$day accuracy and six orders of magnitude less than {that} claimed in \cite{Ciu04}.    Similar conclusions were recently reached in \cite{Nob07} in which LAGEOS and its perigee, along with another similar SLR target were examined. Nobili et al.  write in \cite{Nob07} that  ``it
would need about 120,000 yr of LAGEOS data to get this limit down to the $10^{-12}$
level of EP test already reached by torsion balances!!''.

Alternative orbital configurations for LARES have been recently examined  \cite{Iorio05b,Ciu06}  involving much smaller semimajor axes than that originally proposed.
Let us consider the case of a low-altitude LARES with $a=7,878$ km \cite{Ciu06} (and $i\approx 90$ deg), which was, incidentally, already proven to be unsuitable to measure the Lense-Thirring effect \cite{Iorio07}.
According to \rfr{ciufep}, where, in this case,    $a$ is the semimajor axis of LARES and $d$ is the difference between it and the one of LAGEOS, i.e. $d=4,392$ km, the errors due to $\delta a$ and $\delta d$ are
\begin{equation}
\left\{
\begin{array}{lll}
[\delta(\Delta\psi_N)]_a
\leq
\left|\rp{  -\rp{3d}{2\sqrt{a}}\left(  3d\sqrt{a} + 2\sqrt{a^3} \right) +3d\sqrt{a}\left(\rp{3d}{2\sqrt{a}} +3\sqrt{a} \right) }{\left(  3d\sqrt{a} + 2\sqrt{a^3} \right)^2}\right|\delta a =1\times 10^{-10}\ {\rm cm}^{-1}\delta a ,\\\\

[\delta(\Delta\psi_N)]_d
\leq
\left|\rp{-6a^2}{\left(  3d\sqrt{a} + 2\sqrt{a^3} \right)^2}\right|\delta d=2\times 10^{-11}\ {\rm cm}^{-1}\delta d.
\end{array}\lb{sist}
\right.
\end{equation}
By realistically assuming $\delta a\approx\delta d\approx 1$ cm, \rfr{sist} tells us that we are still three orders of magnitude {above} even the present$-$day level. It is important to note that the results of \rfr{sist} are optimistic: indeed, for the sake of simplicity
we neglected the impact of the first even zonal harmonic $J_2$ which, instead, was considered in \cite{Iorio04} {along with the impact of the Earth's $GM$ as well which can be kept under control for a sufficiently high number of orbital revolutions}.

\section{Conclusions}
In conclusion, after having yielded further arguments concerning the impossibility of using the perigee of LARES/WEBER-SAT for testing the DGP braneworld model, we have shown that, whatever orbital configuration may be devised for the new spacecraft, it will not be able to obtain any improvement in testing the equivalence principle, remaining far from even the present$-$day level of accuracy by three$-$four orders of magnitude.

%


\end{document}